# Dynamic Characterization of Arrows through Stochastic Perturbation

R. Fish, Y. Liang, K. Saleeby, J. Spirnak, M. Sun, X. Zhang

*Abstract*—Current arrow spine measurements rely on statically hanging a known weight at the shaft center and measuring the maximum deflection. This archaic method of measuring arrow stiffness ignores dynamic nature of the arrow when released from the bow. For this project, we built an apparatus to measure the dynamic characteristics of the arrow to better indicate arrow performance. Using stochastic perturbations from a voice coil actuator and displacement measurements, we successfully estimated the natural frequency, damping parameter, and mechanical stiffness of carbon, wood, and aluminum arrows of varying spines. Parameter estimates using a second order parameterized model showed agreement with the manufacturer rated spine values. In addition, high cycle fatigue testing was completed on each arrow material but showed no significant changes in arrow parameters.

*Index Terms*—Arrow, archery, spine, stiffness, stochastic perturbation, system identification.

## I. INTRODUCTION

The bow and arrow, one of the oldest yet most efficient mechanisms developed by early humans, allowed us to hunt game, protect ourselves, and wage war among nations. Much of archery has changed since the traditional long bow. Recurves, compound, and Olympic style bows are all different and are all highly outfitted for their respective purposes. However, the basic principle of storing energy within the bow to rapidly propel an arrow still lies at the heart of all archery.

Despite advancements in bow technology, the design of the arrow has remained largely unchanged. Archers will often spend an exorbitant amount of time and effort tuning individual components on their bow but neglect to optimize the characteristics of the arrows to match the bow. Arrow construction, from front to back, consists of the tip, shaft, and fletching. Characteristics of each include: tip weight, shaft weight/length, shaft spine, and fletching pattern. The combined tip and shaft weight determines the bulk flight characteristics such has distance, speed and target penetration. The shaft spine indicates the arrows stiffness and flex. The fletching induces spin and aerodynamically stabilizes the arrow in-flight. Of these three characteristics, the arrow spine is most difficult to select and must be matched to the bow; weight characteristics can be selected based on the particular application (hunting or target archery) and fletching construction is generally independent of the bow (as long as the individual fletch clears the arrow rest).

The arrow spine directly determines how much flex will be induced in the arrow when shot. For example, a low-stiffness arrow shot from a high-poundage bow will flex more than a high-stiffness arrow shot from a low-poundage bow; both cases are detrimental to the arrow grouping and shot accuracy. In extreme cases, low-stiffness arrows on high-poundage bows can break and cause devastating injuries.

Current arrow spine measurements remain rather archaic. The ASTM F2031-05 standard hangs an 880 g (1.94 lb) weight from the center of a 0.71 m (28 in) section of the arrow shaft and measures the maximum deflection, as shown in Figure 1. The arrow spine is the measured maximum deflection at the center [1]. Using this method, the spine rating is counterintuitive, where higher spine ratings actually indicate a less stiff arrow. In addition, the standard only measures a static displacement, which is indicative but does not fully capture the dynamic performance of the arrow when shot; dynamic parameters such as damping are completely ignored. In this project, we applied system identification principles to dynamically measure the arrow, evaluating stiffness and damping parameters of aluminum, wood, and carbon fiber arrows. In addition, we analyzed the fatigue characteristics of the same arrows to evaluate arrow degradation over time.

## II. EXPERIMENTAL DESIGN

The designed system constrains the arrow at the two ends at a known distance and vibrates the arrow in the center using a Voice Coil Actuator (VCA). Displacement is measured using an integrated sensor in the VCA body. Data acquisition and VCA controls are implemented using a National Instruments myRIO controller and LabVIEW software. Details of each are presented in the following sections.

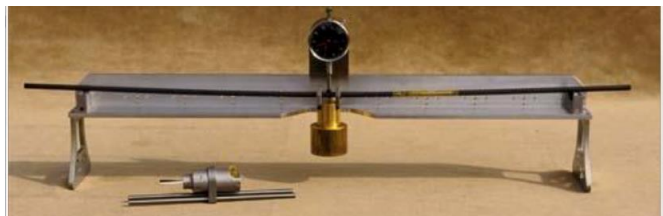
Figure 1: Current arrow spine measurement.

### A. Mechanical Design

After discussing multiple concepts to evaluate the dynamic response of an arrow, we came up with a preliminary design shown in Fig. 2 below. We used a speaker as the driving force to deflect the arrow, and a position sensor to monitor the displacement. An 80-20 rail with end clamps creates a fixed-fixed configuration for the arrow. A 3D printed clamp connects the arrow to the VCA.

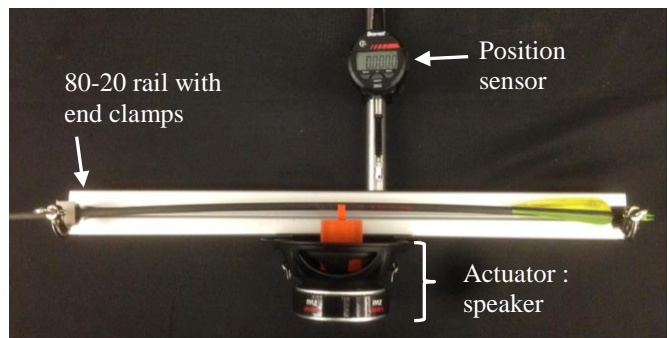

Figure 2: Preliminary design.

This design provided information as to the main frequency range needing testing, and the required rigidity in the arrow mounts. However, there were a few flaws in this setup. One challenge was that the position sensor is not rigidly attached to the arrow during measurement. This diminishes the sensor's responsiveness to the arrow during actuation. Another issue was that the actuator is not rigidly connected to the arrow. This added dynamics between the actuation of the VCA and the arrow.

To solve the first issue, we purchased a VCA (BEI Kimco - LAS16-23-000A-P01-4E)[2] with integrated encoder. The integrated VCA can output 89 N peak force, over 6 mm travel, and has a 10 µm displacement resolution. A 3D printed kinematic clamp was threaded into the VCA output shaft to rigidly connect the actuator and the arrow. The rigid mounting removed unmodeled dynamics.

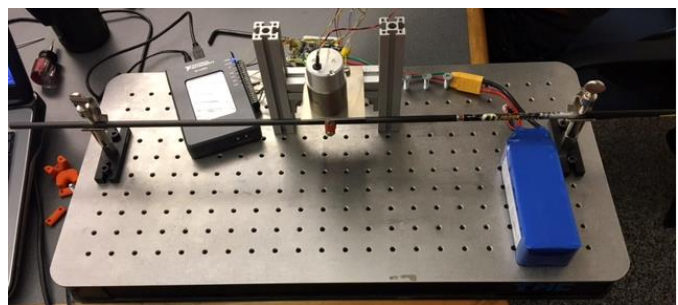

Figure 3: VCA with integrated encoder.

Compared to the preliminary design, the speaker voice coil actuator and the position sensor have been replaced by the VCA in Fig. 3. The 80-20 mounting system has been replaced by an optical mounting board. The clamp in the middle of the arrow completely encases it, providing a tighter grip on the arrow and thus preventing unwanted lateral and rotational movement. The end clamps are mounted on the same mounting board the arrow is mounted on, eliminating any relative movement between the two mounts and thus any noise due to the shifting of the 80-20 during the data acquisition.

### B. Data Acquisition Setup

Informational flow within the system is presented in Fig. 4.

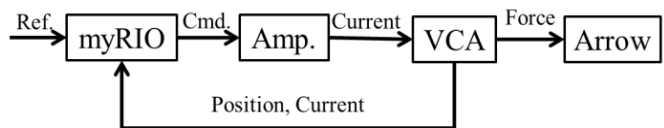

Figure 4: Data acquisition information flow diagram.

Prior to acquisition, desired force commands are generated using MATLAB and sent into the myRIO file storage system. The myRIO then reads and converts the desired force vector into PWM and directional commands compatible with the motor driver (Cytron MD10C). Within each loop iteration, the myRIO updates the desired PWM duty cycle and direction, and reads the position sensor and the sense resistor voltage. The PWM frequency was set to 20 kHz to avoid current ripples and the loop rate was limited to 4 kHz to ensure completion of all tasks within each iteration. The loop rate dictated the maximum sampling frequency at 4 kHz but is sufficient for this particular measurement.

Due to the back EMF reducing the actual current through the VCA, a sense resistor was placed in series with the VCA to measure VCA current. A full close-loop current controller could be implemented to match the output current with the desired force but is outside the scope of this course. The measured sense resistor voltage is converted into force and used as the stochastic input vector for system identification. The measured VCA position is used as the output vector for system identification.

### C. Electrical Design

The electrical design comprises two aspects of the testing rig. The force applied to the arrow is generated electrically through the VCA and the forcing current is measured electrically, and the position of the arrow is measured through an analog output. The circuitry implemented is shown in Fig. 5.

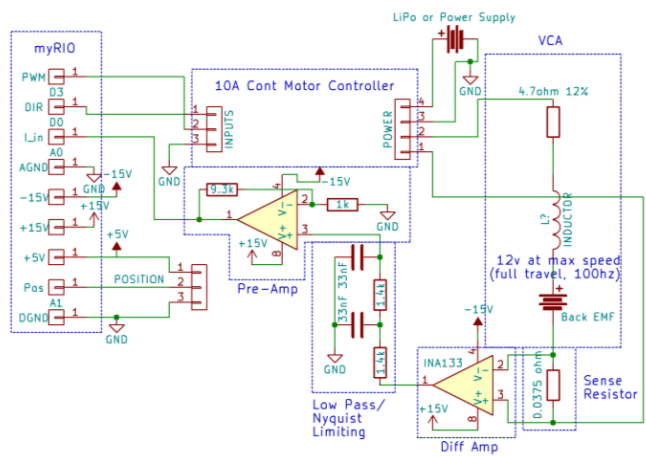

Figure 5: Circuit and filter design

The position output of the VCA was designed by the VCA manufacturer to be directly proportional to the VCA output shaft position relative to the VCA body. This linearity was confirmed observationally during testing. The position measurement was extremely clean, and no further processing was necessary. The measurement interface required a +5 V power and ground reference connection, and provides a 0-5 V analog output of the position signal. The power rails were connected directly to the +5 V supply and ground on the myRIO, and the position signal was connected directly to one of the two ADC inputs available.

The force input to the mechanical domain required greater complexity. Since an aim for the course is to make the testing rig portable, the power electronics should minimize weight and size. A linear power supply, one in which output power is throttled by controlling the voltage dropped in the regulating element, would have required excessively large heatsinking and resulted in poor efficiency, requiring an oversized power source. Since the mechanical domain under consideration is over an order of magnitude slower than the electrical domain, a switched-mode power amplifier was selected.

The information provided with the VCA datasheet suggests that the power-dissipation capabilities of an unmodified VCA limit the current input to 1.5 A continuous, or 7 A for a 10 s pulse from room temperature. The pulse current requires a power supply capable of holding 33 V under high load, and that is under zero back EMF conditions (the mechanical load is static). In order to provide as high current as possible with back EMF measuring near 10 V for full travel motion above 30 Hz, and retain portability, a 6S LiPo hobby-grade battery was chosen for a power supply. This represents a charged voltage of 25.2 V, enough to provide 3 A at high back EMF conditions. A 10 A motor controller capable of 30 V operation was selected to provide voltage, current, and thermal overhead.

Since the design did not perform active current control, or a feed-forward back EMF compensatory control, the force input to the arrow can only be determined through accurate measurement of the current passing through the VCA. To keep the measurement simple, and given the relatively high currents to be measured, a low-resistance sense resistor was placed in series with the VCA. Since the switched-mode power amplifier rapidly switches between applying 25 V and 0 V to a given terminal, a difference amplifier designed for extremely high, 120 dB, common-mode attenuation with over 200 V common-mode input range relative to the supply rails was selected to measure the voltage across the sense resistor. Since the Nyquist frequency is 2000Hz, and significant noise is present at 20 kHz due to the PWM switching, a double-pole RC filter was added with its knee around 2000 Hz to attenuate these noise sources. This filter operates on the output from the difference amplifier, and feeds this result to a final amplifier stage that brings the output voltage up to levels that better suit the input range for the myRIO ADC. The net conversion factor from current through the VCA to voltage into the myRIO was 0.36, selected so that the maximum current through the VCA would result in a 2 V swing at the myRIO.

### D. Measured Arrows

Three commonly used arrow shaft materials were selected to be measured: carbon, aluminum, and wood. For each shaft material, two to three spines were selected. Due to the myriad of naming systems for arrow spines between both manufacturers and materials, all measured arrows were standardized using SI units. Relevant parameters for the measured arrow are summarized in Table 1.

| Material | Deflect. (m) | Stiff. (N/m) | Freq. (Hz) |
|---|---|---|---|
| Carbon (Arr300) | 0.0076 | 1132.9 | 39.4 |
| Carbon (Arr500) | 0.0127 | 679.7 | 36.1 |
| Carbon (Arr600) | 0.0152 | 566.4 | 35.2 |
| Al (Arr2219) | 0.0086 | 1008.5 | 31.9 |
| Al (Arr1916) | 0.0158 | 545.5 | 27.6 |
| Cedar (ArrYel) | 0.0103 | 839.2 | 30.0 |
| Cedar (ArrRed) | 0.0132 | 653.6 | 26.5 |

Table 1: Selected arrows for measurement. Spine values were converted to actual displacement for clarity and stiffnesses and frequencies were calculated using lumped parameter estimates from factory data.

### III. SYSTEM IDENTIFICATION AND ANALYSIS

Stochastic binary input voltage (V) was applied to the voice coil actuator. The intermediate output from the sense resistor (force) and output (displacement) were as shown in Fig. 6. The system that we characterized has intermediate output force as its input, and displacement as its output.

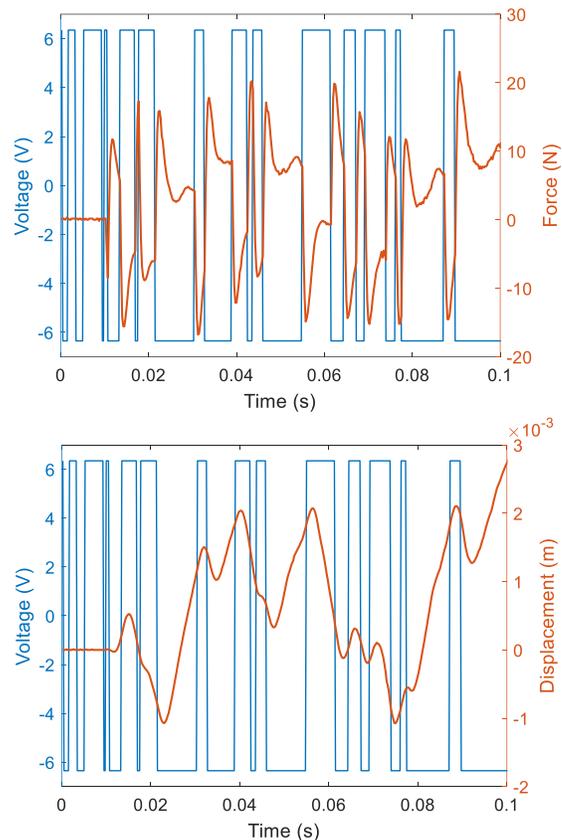

Figure 6: (Top) Stochastic binary input voltage (V) and intermediate output force (N) versus time (s). (Bottom) Stochastic binary input voltage (V) and output displacement (m) versus time (s).

### A. Non-Parametric Model

To determine the non-parametric impulse response, the auto-correlation of the input force was deconvolved from the cross-correlation of the input force and output displacement using a Toeplitz matrix inversion technique. The impulse response for Arr300 for one trial is indicated in Fig. 7. By convolving the input force with the impulse response, the non-parametric predicted output is shown in Figure 8, with the output variance accounted for (VAF) equal 99.42%.

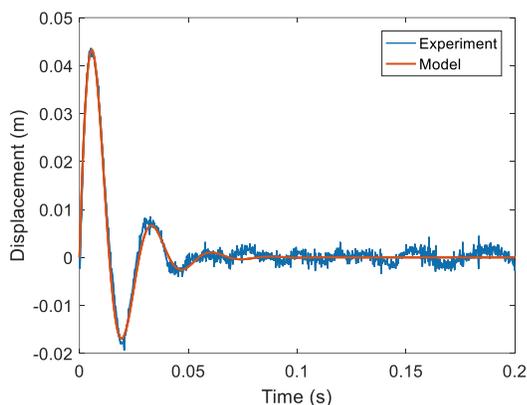

Figure 7: Impulse response. Impulse response acquired from experiment (blue); impulse response acquired by fitting using a second-order system model with no zero (orange).

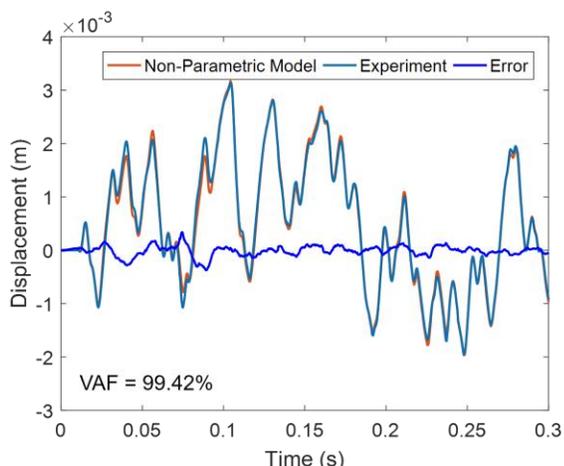

Figure 8: Comparison of the predicted output displacement (m) by non-parametric model and the output displacement (m) from one iteration of experiments for Arr300. The output variance accounted for (VAF) is 99.42%.

*B. Parametric Model*

The governing equation for the dynamics of the system is the Euler-Lagrange equation for beams, which gives a second derivative of the displacement with respect to time and a fourth derivative with respect to position. Therefore, a reasonable parametric mathematical model for the system may start from a non-zero second-order transfer function between the displacement and the force, representing the Kelvin-Voigt model of viscoelastic materials. The parametrized model time-domain equations are shown in the Appendix.

The model in Eq. 1 is fit through the method of least squares using Levenberg-Marquadt nonlinear minimization. There are three parameters in this model, which is the DC gain, the natural frequency and damping parameter. The fitting parameters are listed in Table 2. This 3 parameter linear dynamic model gives the output variance accounted for (VAF) equal 98.17%, demonstrating a high degree of agreement between the model and the actual system. The discrepancy may be due to a combination of noise and modeling error, such as non-linearity.

As noted, the VAF by the non-parametric model will always be greater than the VAF by the parametric model, because the parametric model has 3 parameters to fit whereas the non-parametric model essentially has 1501 parameters in our experiments.

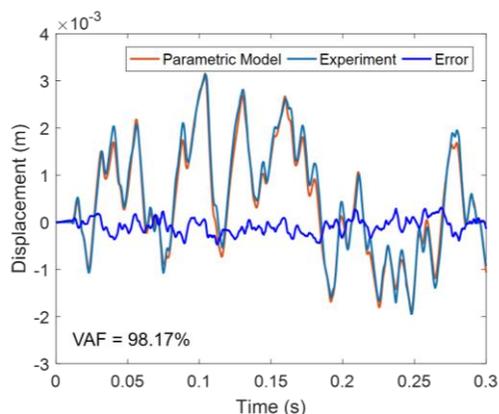

Figure 9: Comparison of the predicted output displacement (m) by parametric model of a second-order system with no zero and the output displacement (m) from one iteration of experiment for Arr300. The output variance accounted for (VAF) is 98.17%.

From the perspective of step response, a zero in the left half-plane (LHP) in *s*-domain will increase the overshoot if the zero is within a factor of 4 of the real part of the complex poles; a zero in the RHP will depress the overshoot and may cause the step response to start out in the opposite direction to the input signal. As shown in Table 2, the parametric model fitting of Eq. 2 for Arr300 gives an LHP-zero, indicating a minimum-phase system. The output variance accounted for (VAF) is 98.29%, which results in a slight improvement.

Given anti-resonance at specific frequencies produced by the simplified Euler-Bernoulli beam model for the arrow, a pair of conjugate zeros are added in the parametric mathematical model for the system. The parametric model fitting of Eq. 3 for Arr300 gives us a pair of zeros with a larger resonant frequency than the natural frequency of the modelled system. The locations of poles and zeroes and its Bode diagram are plotted in Fig. 10. The experiment result is shown in Table 2. No significant improvement for VAF has been noticed.

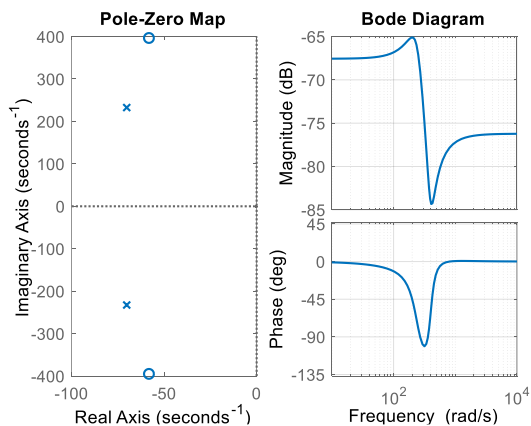

Figure 10: Locations of poles and zeros and Bode diagram for the model of a second-order system with a pair of zeros.

Therefore, parametric model of a second-order system with no zero is applied for all types of arrows. The parametric models for all types of arrows are shown in Table 2.

|  | Gain | Dmp. | Freq. (rad/s) | Zero Pos. | VAF (%) |
|---|---|---|---|---|---|
| 2nd-order model | 2.64e-4 | 0.285 | 239.16 | N/A | 98.17 |
| 2nd-order one zero | 2.64e-4 | 0.289 | 243.01 | z = -4163 | 98.29 |
| 2nd-order pair of zeros | 4.18e-4 | 0.289 | 243.01 | z = -6.48 +/-21.40$i$ | 98.30 |
| Non-parametric model |  |  |  |  | 99.42 |

Table 2: Comparison between different parametric models for Arr300

## IV. RESULTS AND DISCUSSION

### A. Parameter Estimates

From the system identification, the following parameters were estimated for each arrow: resonant frequency, damping parameter, lumped mass, damping value, and stiffness value. Cedar arrows were measured with wood grains both vertical and horizontal. Averaged estimates and max/min variation for each arrow set is presented in Table 3 below.

| Material | Freq. (Hz) | Dmp. Rat. | M (g) | B (Ns/m) | K (N/m) |
|---|---|---|---|---|---|
| Carbon (300) | 40.3 ± 1.1 | 0.32 ± 0.005 | 13.5 ± 0.4 | 2.2 ± 0.5 | 853.6 ± 175.3 |
| Carbon (500) | 34.5 ± 1.9 | 0.39 ± 0.014 | 10.3 ± 0.0 | 1.7 ± 0.1 | 486.2 ± 67.4 |
| Carbon (600) | 35.3 ± 1.2 | 0.51 ± 0.026 | 9.8 ± 0.2 | 2.3 ± 0.5 | 490.8 ± 109.7 |
| Al (2219) | 40.8 ± 1.8 | 0.67 ± 0.015 | 22.7 ± 0.2 | 7.8 ± 0.5 | 1484.7 ± 29.9 |
| Al (1916) | 31.6 ± 1.5 | 0.53 ± 0.163 | 14.8 ± 0.2 | 3.1 ± 0.7 | 585.1 ± 109.9 |
| Cedar (405V) | 38.3 ± 0.2 | 0.52 ± 0.048 | 15.4 ± 0.0 | 3.7 ± 0.4 | 891.6 ± 4.3 |
| Cedar (405H) | 37.4 ± 0.5 | 0.53 ± 0.075 | 17.5 ± 0.2 | 4.5 ± 0.5 | 967.7 ± 131.7 |
| Cedar (520V) | 31.9 ± 0.7 | 0.50 ± 0.052 | 14.7 ± 0.1 | 3.1 ± 0.6 | 592.2 ± 81.7 |
| Cedar (520H) | 31.3 ± 0.7 | 0.55 ± 0.008 | 13.6 ± 0.1 | 2.9 ± 0.5 | 525.5 ± 3.2 |

Table 3: Compiled parameter estimates for all measured arrows. Maximum and minimum ranges were calculated for each arrow group.

In comparison to the expected resonance frequencies presented in Table 1, parameter estimates from the system identification shows positive agreement with the expected values. From the system identification, different arrow spines can be clearly distinguished by the resonance frequency. With the exception of the carbon 500 and 600, the general trend of decreasing stiffness for higher rated spine exists for all arrow materials. Damping parameter vary widely within each arrow material. Carbon arrows have the least amount of damping ranging between 0.3 and 0.5 while aluminum arrows showed high damping between 0.5 and 0.7. Wood arrows stayed consistently ~0.5 without noticeable change depending on the grain orientation.

### B. Sensitivity

Sensitivity analysis was conducted for the 3rd order parameterized model. Each of the three parameters were varied (DC gain, damping parameter, and natural frequency) while two were held fixed. VAF was observed to determine the necessity and effect of each parameter on the model fit. VAF was significantly affected by each parameter, demonstrating that each parameter in our 3rd order model was necessary.

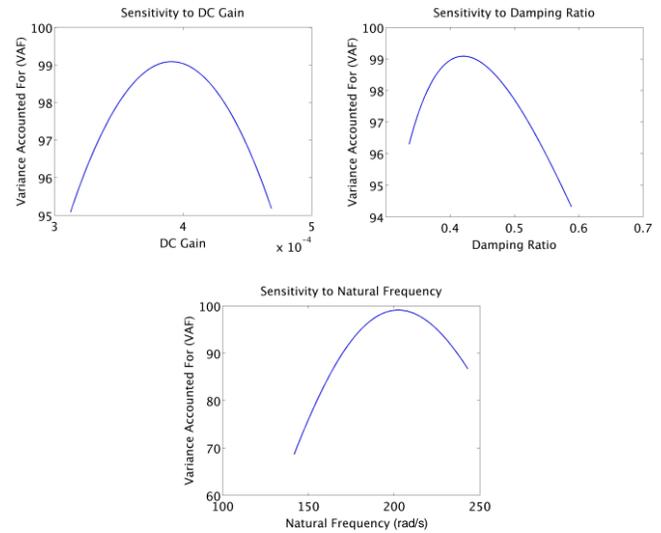

Figure 11: Sensitivity analysis of parameterized model evaluated based on Variance Accounted For.

### C. Fatigue

One arrow for each arrow material was selected for fatigue testing. Each arrow was fatigued for ~600,000 cycles and measured on the system again. Estimated parameters from the system identification are summarized in Table 4 below.

| Material | Freq. (Hz) | Dmp. Rat. | M (g) | B (Ns/m) | K (N/m) |
|---|---|---|---|---|---|
| Carbon (500) | 45.5 | 0.42 | 9.024274 | 2.15 | 737.2 |
| Al (2219) | 43.8 | 0.35 | 22.55158 | 4.35 | 1705.4 |
| Cedar (405V) | 40.2 | 0.51 | 15.95976 | 4.15 | 1019.4 |
| Cedar (405H) | 41.4 | 0.50 | 18.52453 | 4.83 | 1251.0 |

Table 4: Parameter estimates for arrows after ~600,000 high cycle fatigue.

600,000 cycles is more cycles than a common arrow would endure over its lifetime. Estimating 10 cycles per shot, 30 shots per day, 600,000 cycles would last over 5 years, well higher than any arrow would endure unless it is shared in a club shooting setting. Based on the parameter estimates, there was no significant change in the fatigued arrows. Slight changes in damping and increase in stiffness can be seen but a reasonable conclusion cannot be drawn due to the small sample size

## V. CONCLUSION AND FUTURE WORK

The designed apparatus was successfully able to estimate bulk arrow parameters using a stochastic perturbation and a second order model. As expected, arrows with lower spines showed higher stiffness values. However, measurements for the same arrow spine varied greatly between arrows. This may indicate: (1) manufacturing variation between arrows, (2) inconsistent end clamping between arrow measurement, or (3) errors within the parameter estimates. A full calibration of the apparatus with standards of known stiffnesses is required to identify the root of the issue but is outside the scope of this course. In addition to the stiffness, estimated parameters showed variations in damping parameter between different arrow materials. Carbon arrows showed the lowest damping parameter, ~0.3, while wood and aluminum showed ~0.5. Wood grain orientation showed no effect on the stiffness nor the damping parameter. High cycle fatigue of all arrow materials also showed no significant change in any of the parameters, indicating that an arrow will likely break due to normal wear before high cycle fatigue failure. To generalize these parameter estimates and indicated trends, a larger sample of arrows is needed. The system does successfully demonstrate the feasibility of using stochastic perturbations to identify dynamic characteristics of an arrow and provide additional information beyond the current hanging weight standard. With a calibrated measurement system, it would be interesting to evaluate how arrow damping actually affects archery performance.

## Appendix

Eq. 1: $y(t) = G \frac{\omega}{\sqrt{1-\zeta^2}} e^{-\zeta\omega t} \sin(\omega t \sqrt{1-\zeta^2})$

Eq. 2: $y(t) = G\omega^2 e^{-\zeta\omega t} \cosh(\omega t \sqrt{1-\zeta^2}) + \frac{z-\zeta\omega}{\omega\sqrt{1-\zeta^2}} \sinh(\omega t \sqrt{1-\zeta^2})$

Eq. 3: $y(t) = \frac{G\omega^2}{z^2} \delta(t) - \frac{1}{z^2} e^{-\zeta\omega t} \cosh(\sqrt{1-\zeta^2}) + \frac{2G\omega^2(\omega-\xi z)}{\omega z^4 \sqrt{1-\zeta^2}} e^{-\zeta\omega t} \sinh(\omega t \sqrt{1-\zeta^2}) \left[\frac{\omega^2-z^2}{2(\zeta\omega-\xi z)} - \zeta\omega\right]$